\documentclass[article]{aa}  
\usepackage{rotating}
\usepackage{graphicx}
\usepackage{tikz,hyperref}
\usepackage[capitalize]{cleveref}
\defcitealias{Hovatta2012}{H12}
\crefformat{section}{Sect. #2#1#3}

\definecolor{lime}{HTML}{A6CE39}
\DeclareRobustCommand{\orcidicon}{%
    \begin{tikzpicture}
    \draw[lime, fill=lime] (0,0) 
    circle [radius=0.16] 
    node[white] {{\fontfamily{qag}\selectfont \tiny ID}};
    \draw[white, fill=white] (-0.0625,0.095) 
    circle [radius=0.007];
    \end{tikzpicture}
    \hspace{-2mm}
}

\hypersetup{
    pdftitle={The Faraday Rotation of NRAO\,150},
    pdfauthor={Jack Livingston},     
    colorlinks=true,       
    linkcolor=purple,
    filecolor=magenta,      
    urlcolor=violet,
    citecolor=teal,        
    pdfborder={0 0 0}
}
\newcommand{\orcidJDL}{\href{https://orcid.org/0000-0002-4090-8000}{\orcidicon}}
\newcommand{\orcidYYK}{\href{https://orcid.org/0000-0001-9303-3263}{\orcidicon}}
\newcommand{\orcidMW}{\href{https://orcid.org/0000-0002-8635-4242}{\orcidicon}}
\newcommand{\orcidGFP}{\href{https://orcid.org/0000-0001-6757-3098}{\orcidicon}}
\newcommand{\orcidSAD}{\href{https://orcid.org/0000-0001-6010-6200}{\orcidicon}}
\newcommand{\orcidMML}{\href{https://orcid.org/0000-0001-6088-3819}{\orcidicon}}
\newcommand{\orcidASK}{\href{https://orcid.org/0000-0002-4009-9186}{\orcidicon}}
\newcommand{\orcidJR}{\href{https://orcid.org/0000-0002-2426-927X}{\orcidicon}}
\newcommand{\orcidNRM}{\href{https://orcid.org/0000-0002-6684-8691}{\orcidicon}}
\newcommand{\orcidLCD}{\href{https://orcid.org/0009-0003-8342-4561}{\orcidicon}}
\newcommand{\radm}{$\mathrm{\,rad\,m^{-2}}\,$}

\usepackage{txfonts}
\titlerunning{The Faraday Rotation of NRAO\,150} 
\authorrunning{J. D. Livingston et al.}
\begin{document}

\title{A helical magnetic field in quasar NRAO\,150\\ revealed by Faraday rotation}

   \author{J. D. Livingston\orcidJDL\inst{1}\fnmsep\thanks{jack.david.livingston+academic@gmail.com},
    A. S. Nikonov\orcidASK \inst{1};
          S. A. Dzib\orcidSAD \inst{1},
          L. C. Debbrecht\orcidLCD \inst{1},
          Y. Y. Kovalev\orcidYYK\inst{1},  \\
          M. M. Lisakov\orcidMML \inst{1,} \inst{2,} \inst{3},
           N. R. MacDonald\orcidNRM \inst{4}, 
          G. F. Paraschos\orcidGFP \inst{1}, 
          J. Röder\orcidJR \inst{1}, and
          M. Wielgus\orcidMW \inst{5,} \inst{1}
          }

   \institute{\inst{1} Max Planck Institute for Radio Astronomy (MPIfR),
              Auf dem Hügel 69, 53121 Bonn\\
              \inst{2} Lebedev Physical Institute of the Russian Academy of Sciences, Leninsky prospekt 53, 119991 Moscow, Russia \\
              \inst{3} Instituto de Física, Pontificia Universidad Católica de Valparaíso, Casilla 4059, Valparaíso, Chile \\
              \inst{4} The University of Mississippi, Department of Physics and Astronomy, Oxford, US \\
              \inst{5} Instituto de Astrofísica de Andalucía-CSIC, Glorieta de la Astronomía s/n, E-18008 Granada, Spain \\
              \email{jack.david.livingston+academic@gmail.com}}

   \date{Received XX; accepted YY}

 
  \abstract
   {Active Galactic Nuclei (AGN) are some of the most luminous and extreme environments in the Universe. The central engines of AGN, believed to be super-massive black-holes, are fed by accretion discs threaded by magnetic fields within a dense magneto-ionic medium.}
   {We report our findings from polarimetric Very-long-baseline Interferometry (VLBI) observations of quasar NRAO\,150 taken in October 2022 using a combined network of the Very Long Baseline Array (VLBA) and Effelsberg 100-m Radio Telescope. These observations are the first co-temporal multi-frequency polarimetric VLBI observations of NRAO\,150 at frequencies above 15\,GHz.}
   {We use the new VLBI polarization calibration procedure, \texttt{GPCAL}, with polarization observations of frequencies of 12\,GHz, 15\,GHz, 24\,GHz, and 43\,GHz of NRAO\,150. From these observations, we measure Faraday rotation. Using our measurement of Faraday rotation, we also derive the intrinsic electric vector position angle ($\mathrm{EVPA_0}$) for the source. As a complementary measurement we determine the behavior of polarization as a function of observed frequency.}
   {The polarization from NRAO\,150 only comes from the core region, with a peak polarization intensity occurring at 24\,GHz. Across the core region of NRAO\,150 we see clear gradients in Faraday rotation and $\mathrm{EVPA_0}$ values that are aligned with the direction of the jet curving around the core region. We find that for the majority of the polarized region the polarization fraction is greater at higher frequencies, with intrinsic polarization fractions in the core $\approx 3\%$. }
   {The Faraday rotation gradients and circular patterns in $\mathrm{EVPA_0}$ are strong evidence for a helical/toroidal magnetic field, and the presence of low intrinsic polarization fractions indicate that the polarized emission and hence the helical/toroidal magnetic field, occur within the innermost jet.}

   \keywords{quasars: individual: NRAO\,150 --
                Magnetic fields --
                Polarization
               }

   \maketitle
%

\section{Introduction}

NRAO\,150 is a highly variable radio loud quasar with a redshift of $z=1.52$ \citep{AcostaPulido2010} and reported viewing angles of $\theta \approx 8^{\circ}$ \citep{Agudo2007} and $\theta \approx 0.28^{\circ}$ \citep{Homan2021}.  The source has an apparent jet opening angle of $19.4^{\circ}$ \citep{Pushkarev2017}, maximum recorded jet speed of $\beta = 8.65$ \citep{Homan2021}, based on the maximum measured jet speed over 42 epochs at 15\,GHz, and a variability Doppler factor $\delta_{\mathrm{var}} =  13.1$ \citep{Hovatta2009}. \cite{Agudo2007} found that the inner jet of NRAO\,150 swings $\approx 6^{\circ} - 11^{\circ}$ on a timescale of one year using five epochs at 43 and 86\,GHz.

The polarization properties of NRAO\,150 were studied by \citet{Molina2014} using observations at 8, 15, 22, 43, and 86\,GHz observed using the Very Long Baseline Array (VLBA) and Global Millimeter VLBI Array (GMVA) from 2006 to 2009, and a follow-up study was conducted using additional 2008 -- 2009 observations at 22, 43, and 86\,GHz with the VLBA and GMVA \citep{Molina2016}. These studies found that the electric vector polarization angle (EVPA) of the source showed a circular pattern which they concluded was evidence of a magnetic field with helical/toroidal geometry. 

\citet{Paraschos2024} used European VLBI Network (EVN) polarimetric data of NRAO\,150 at 22\,GHz to calibrate their 3C\,84 data. These NRAO\,150 data will be present in Debbrecht et al. in prep. The EVPA they found shows a circular pattern in the core (similar to that found by \citealt{Molina2016}) and EVPA values aligned with the direction of the bulk jet flow further downstream. Within the core they measured polarization intensities peaking at $\sim 23$ mJy/beam. 

When polarized emission passes through a magneto-ionic medium, the EVPA is expected to rotate via birefringence. This effect is called "Faraday rotation" and is proportional to the square of the observed wavelength. Typically, we assume that this medium or "screen" is external to the polarized emission region. We quantify the amount of Faraday rotation as the observed rotation measure, $\mathrm{RM_{obs}}$, which we characterize as $\mathrm{EVPA} = \mathrm{EVPA_0} + \mathrm{RM_{obs}} \lambda^2$. Here, $\mathrm{EVPA_0}$ is the Faraday corrected or "intrinsic" EVPA at the source. 

RM relates to the electron number density, $n_e$, in units of $\mathrm{cm^{-3}}$, and the line-of-sight (LOS) magnetic field, $B_{||}$ as, 
\begin{equation}
\mathrm{RM} \equiv \mathcal{C} \int_{r=0}^{r=d} n_e\,B_{||} d\,\textbf{\textit{r}}\, (\mathrm{rad\,m^{-2}}). 
\label{eqn:RM}
\end{equation}
Here, $d$ is the distance to the Faraday rotating medium in pc, \textit{d}\textbf{\textit{r}} is the incremental displacement along the LOS measured in pc from the observer to the source, and $\mathcal{C}$ is a conversion constant, $\mathcal{C} = 0.812 \, \mathrm{rad\,m^{-2}\,pc^{-1}\,cm^{3}\,\mu G^{-1}}$ \citep{Ferriere2021}. We adopt the vector orientation and sign convention for Faraday rotation as outlined by \cite{Ferriere2021}.

The Faraday rotation of NRAO\,150 was previously measured by \citet{Zavala2004}, observed in 2001, and in a follow-up analysis of the data by \citet{Gabuzda2015}. \citet{Zavala2004} found the majority of Faraday rotation to occur near the optically thick core region and found hints of Faraday rotation that appeared to break the $\lambda^2$ relation. \citet{Gabuzda2015} re-examined the data from \citet{Zavala2004} and found a prominent gradient in Faraday rotation across the core region with a significance of $4.2 \sigma$; they concluded such a significant gradient was evidence for a helical/toroidal magnetic field. NRAO\,150 is a highly variable source; yet, since the 2001 observation from \citet{Zavala2004}, there has been no new work on the Faraday rotation of NRAO\,150. Our work presents the first new Faraday rotation analysis for NRAO\,150 in two decades, and also offers a higher frequency range than previously covered, allowing us to better understand the magnetic fields of the source closer to the central engine.

For this work we assume $H_0 = 71\,\mathrm{km\, s^{-1}\, Mpc^{-1}}$, $\Omega_m = 0.27$, and $\Omega_{\Lambda} = 0.73$ \citep{Komatsu2009}. This corresponds to a luminosity distance for NRAO\,150 of 11\,193.4\,Mpc, with a projected angular scale of 8.55 pc\,mas$^{-1}$. Due to the variation in the viewing angles for this source ($0.28^{\circ}$ or $8^{\circ}$)\footnote{For $\theta \approx 0.28^{\circ}$ the de-projected scale is 1749 pc\,mas$^{-1}$, whereas $\theta \approx 8^{\circ}$ results in a scale of 61 pc\,mas$^{-1}$.} we keep all physical distances as projected distances.

\section{Observations and methods}
NRAO\,150 was observed on October 1, 2022, using the VLBA and the Effelsberg 100m Radio Telescope (EF) as a joint VLBI network, at central frequencies of 12.168, 15.368, 23.568, and 43.169\,GHz, each split into eight intermediate frequencies or IF windows. This observation also included the sources 3C\,111, 3C\,120, and NGC\,1052, as well as a complementary observation of 2023+335, 3C\,371, 3C\,418, BL\,Lac and Cygnus-A on October\,2, 2022. Eight out of ten VLBA antennas were available for this observation (BR, FD, LA, NL, OV, SC, MK, and HN).

\subsection{A-priori and self-calibration}
\label{sec:apriori}
As NRAO\,150 is a very bright and compact AGN with typical VLBI peak flux densities exceeding 5\,Jy\,beam$^{-1}$, it was used as a calibrator source in a-priori calibration using the standard procedures described in the AIPS cookbook\footnote{\href{http://www.aips.nrao.edu}{http://www.aips.nrao.edu}} to test for this flux density loss.

After a-priori calibration, we found a lower overall flux density for all sources within our observations compared to MOJAVE at 15 GHz \citep{Lister2018} and Boston BEAM-ME 43\,GHz \citep{Jorstad2016} measurements of shared sources. We also performed a-priori calibration following using the \texttt{rPicard/CASA  v7.8.1}\footnote{\href{https://bitbucket.org/M_Janssen/picard/src/master/}{https://bitbucket.org/M\_Janssen/picard/src/master/}} pipeline \citep{Janssen2019}. We found that the AIPS a-priori calibrated data shows an overall higher amplitude by 9\% than the data processed with rPicard/CASA. A flux density difference of $\sim10$\% has been seen in preliminary analysis conducted by the MOJAVE team comparing rPicard/CASA and AIPS and is currently being investigated. It is worth noting that according to the NRAO flux density calibration with the VLBA is only good to $\sim 10$\%\footnote{\href{https://science.nrao.edu/facilities/vlba/docs/manuals/propvlba/calibration-considerations}{https://science.nrao.edu/facilities/vlba/docs/manuals/propvlba/calibration-considerations}}, as such the difference of 9\% we see is within that margin. Technical issues as well as weather present challenges for accurate VLBI amplitude calibration. Overall amplitude depends on how bandpass solutions are re-normalized affecting the final values (VLBA Scientific Memo 40\footnote{\href{https://library.nrao.edu/public/memos/vlba/sci/VLBAS_40.pdf}{https://library.nrao.edu/public/memos/vlba/sci/VLBAS\_40.pdf}}). For this reason, MOJAVE uses OVRO \citep{Richards2011} and BEAM-ME uses Metsahovi single dish data\footnote{\href{https://www.metsahovi.fi/opendata/}{https://www.metsahovi.fi/opendata/}} to calibrate the flux density scale.

Even accounting for this flux density uncertainty we find additional losses in both the AIPS and rPicard/CASA calibrated data as compared to MOJAVE and BEAM-ME data. This flux density loss comes from erroneous correlation of the VLBA+EF network which is independent of the flux density difference seen between AIPS and rPicard/CASA pipelines. This error occurred due to a mismatch between the EF and VLBA frequency setups which was not accounted for in correlation. This has been confirmed with the operator of EF. This effect remains if we completely flag EF, as such we require a correction to our absolute amplitude based on external data.

We have elected to correct our AIPS calibrated data, due to better overall dynamic range. We also discovered an issue with rPicard/CASA polarization calibration which produces a $45^{\circ}$ EVPA offset in our observation due to flipping Stokes Q and U. To correct for overall reduction due to erroneous correlation of the VLBA+EF network, we used TPOL0003 VLA observations of NRAO\,150 using the Very Large Array (VLA) conducted on October 10, 2022, across similar frequency bands as our data. We calibrated these observations using the standard VLA \texttt{CASA} calibration pipeline. We then generated images of the calibrated data of NRAO\,150, and compared the derived flux density of the VLA images with the flux density of our VLBA+EF images, finding the multiplicative gain factor to correct for. These gain corrections are shown in \cref{tab:calib}. 

We performed amplitude and phase self-calibration on NRAO\,150 for all frequencies down to 30\,s solution intervals. The gains of EF were considerably lower than expected for the 15 and 24 GHz bands, which was seen for all sources within our observation. As such we used the VLBA antennas to anchor the gains of EF before proceeding with amplitude self-calibration to preserve the absolute amplitude of the source. 

\subsection{Polarization calibration}
\label{sec:pol}
Between a-priori and absolute EVPA calibration, we used \texttt{GPCAL}\footnote{\href{https://github.com/jhparkastro/gpcal}{https://github.com/jhparkastro/gpcal}} \citep{Park2021} to solve for polarization leakage or "D-terms". For leakage calibration we selected NRAO\,150, 3C\,111, and 3C\,120 as calibrators, using an iterative self-calibration approach for ten repeated iterations. 

Absolute EVPA calibration was done by "anchoring" to an external observation. We used the TPOL0003 VLA observations of NRAO\,150. Polarization calibration was done following the methods outlined in the VLA \texttt{CASA} polarization calibration tutorial\footnote{\href{https://casaguides.nrao.edu/index.php/CASA_Guides:Polarization_Calibration_based_on_CASA_pipeline_standard_reduction:_The_radio_galaxy_3C75-CASA6.5.4}{https://casaguides.nrao.edu/index.php/CASA\_Guides}}. We created a point source model of EVPA for the source from 12 to 43\,GHz and calculated the EVPA offset between the VLA and the integrated EVPA from our D-term corrected VLBA+EF data of NRAO\,150 data for each IF window of each band. 

\begin{figure*}
    \centering
    \includegraphics[width=2\columnwidth]{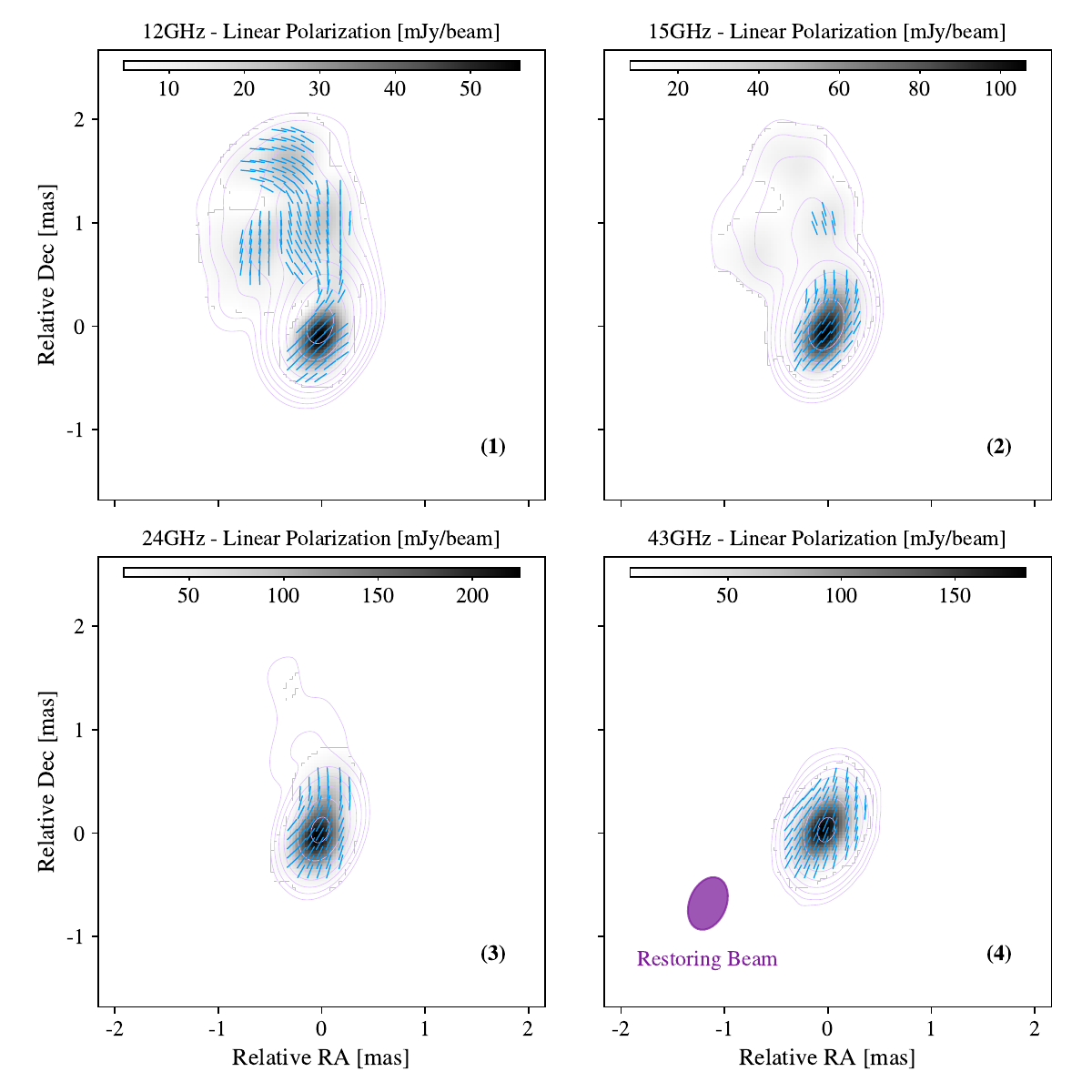}
    \caption{Panels {(1)} to {(4)} show the Stokes I contours and linear polarization intensity (shown in gray-scale) of NRAO\,150 for 12 {(1)}, 15 {(2)}, 24 {(3)}, and 43\,GHz {(4)}. The purple ellipse in panel {(4)} shows the restoring beam used for all four frequencies. The contours start at $5 \sigma_{I}$ up to the peak increasing by a factor of 2. The blue ticks in panels {(1)} to {(4)} show the average $\mathrm{EVPA}$ across a $5 \times 5$ pixel region for each frequency.}
    \label{fig:stki}
\end{figure*}

\subsection{Imaging and error estimation}

\begin{figure*}
    \centering
    \includegraphics[width=2\columnwidth]{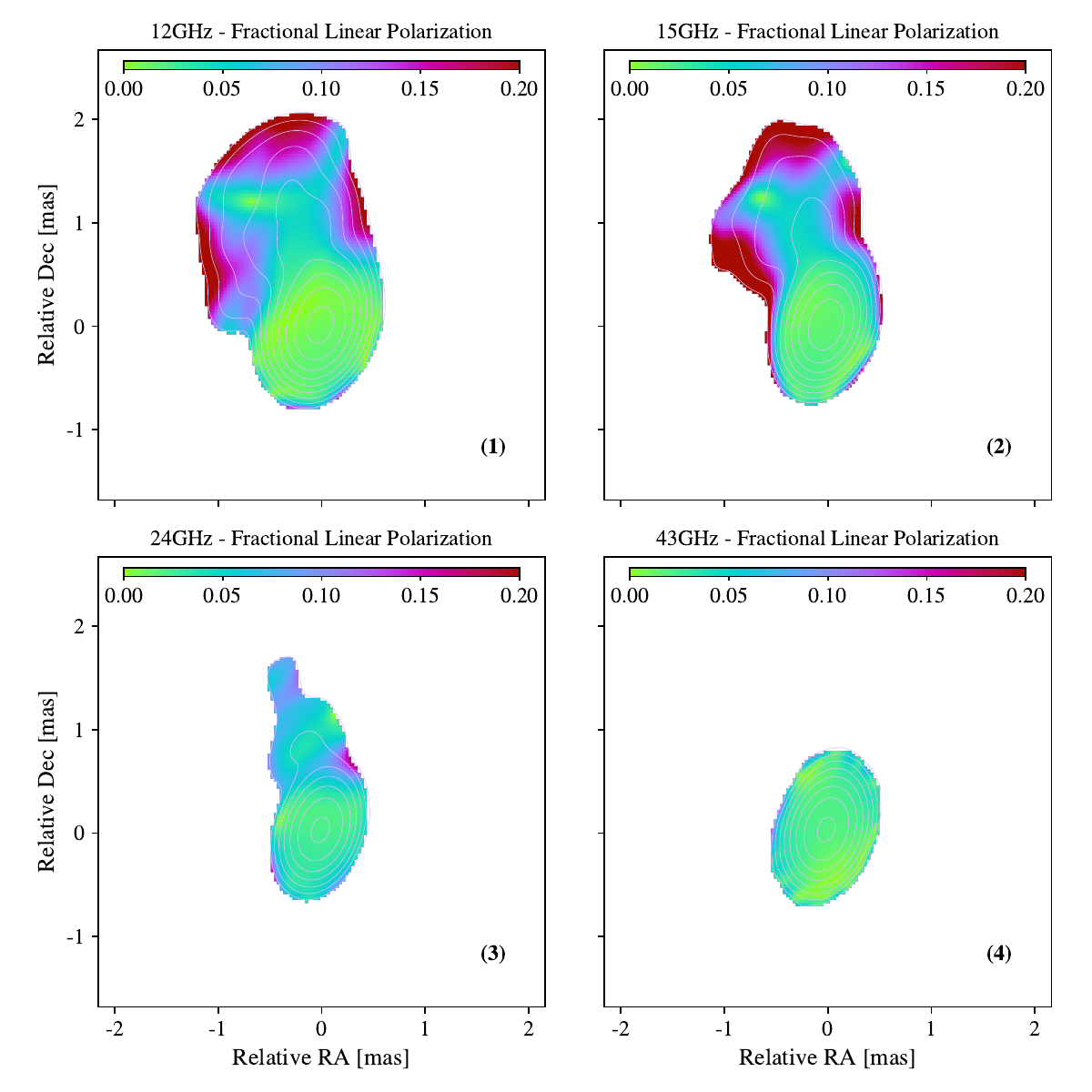}
    \caption{Panels {(1)} to {(4)} show the fractional linear polarization for each frequency. The contours are the lowest Stokes I contour levels for all frequencies from \protect\cref{fig:stki}.}
    \label{fig:frac}
\end{figure*}

The images of Stokes I, Q, and U were generated using the entropy stopping criteria imaging script as outlined by \citet{Homan2024}. The script is based on the VLBI imaging software \texttt{Difmap} \citep{Shepherd1997}, and images the data using a variety of visibility weightings - specifically, 10 (ultra-uniform), 2 (uniform), and 0 (natural), to catch any residual structure left in the data.
To generate the images shown in this work, we used a common resolving element ("restoring beam") across all frequencies with a major axis of $0.53\, \mathrm{mas}$, a minor axis of $0.35\, \mathrm{mas}$, and a position angle of $-22.4^\circ$. In \cref{fig:stki} we present the generated linear polarization images and Stokes I contours for each observed frequency.

We estimated the Stokes I uncertainty $\sigma_{I}$ by finding the root-mean-squared (RMS) noise of a $300\times300$ pixel region sufficiently far away from Stokes I emission of the source. We further estimated the error introduced from residual D-terms ($\sigma_{\mathrm{dterm}}$) to the polarized visibilities after calibration using the same approach as shown in Eq. B4 in \citet{Hovatta2012}, finding the median scatter of D-terms for each frequency band, and accounting for six scans of NRAO\,150, eight spectral windows, and between seven\footnote{The FD antenna was not present for the 43\,GHz observation.} to eight antennas. The scatter of D-terms are shown in \cref{tab:calib}. 

The uncertainty of the polarized signal was determined by finding the RMS of a $300\times300$ square pixel area away from the source of the Stokes Q and U images ($\sigma_{\mathrm{rms}}$) and combining this uncertainty estimate with that from $\sigma_{\mathrm{dterm}}$ following equation B5 of \citet{Hovatta2012}. The uncertainty of the polarization data ($\sigma_p$) is the combination of the uncertainties from Stokes Q and U such that $\sigma_P = \sqrt{\sigma_Q^2 + \sigma_U^2}$.

To estimate the uncertainty of our EVPA measurements, we take the uncertainty assumed from $\sigma_P$, which is ${\sigma_{\mathrm{EVPA}} = \sigma_P/(2 P)}$ and add an additional error based on the scatter of the VLA data around the model of EVPA used for absolute EVPA calibration. This scatter of EVPA values are shown in \cref{tab:calib}.

For image alignment between observed frequencies we implemented a 2D cross-correlation approach on optically thin Stokes I emission similar to that outlined by \citet{Molina2014}. We use the Python package \texttt{phase\_cross\_correlation} from \texttt{skimage.registration}\footnote{\href{https://scikit-image.org/docs/stable/api/skimage.registration.html}{https://scikit-image.org/docs/stable/api/skimage.registration.html}}, but we found that the overall displacement is consistent with zero across all frequencies. \citet{Hovatta2012} showed that the measurement of $\mathrm{RM}$ is robust against a small misalignment between frequencies up to 0.15\,mas for the VLBA at 15\,GHz, as such we expect image alignment errors to not affect our work.

\begin{table*}
    \begin{center}
    \setlength{\tabcolsep}{0.435em}
    \renewcommand{\arraystretch}{1.2}
    \caption{Observed bands, values used through data calibration, and peak Stokes I and linear polarization intensities. }\label{tab:calib}
    \begin{tabular}{c|c|c|c|c|c|c|c|c|c|c|c|c|c}
    \hline\hline
    Band   & $\nu_c$& Range & $b_{\mathrm{maj}}$ & $b_{\mathrm{min}}$ & $b_{\mathrm{PA}}$ &$G_c$  & $\sigma_{\rm dterms}$ & EVPA$_{\rm VLA}$ & $\sigma_{\rm EVPA}$  & Peak I  & $\sigma_{I}$ & Peak P & $\sigma_{P}$ \\
    name   & [GHz] & [GHz]  &  [mas]  & [mas] & [$^{\circ}$] &    &       & [$^{\circ}$]  & [$^{\circ}$]  &  [$\mathrm{\frac{mJy}{beam}}$] &    [$\mathrm{\frac{mJy}{beam}}$]   &   [$\mathrm{\frac{mJy}{beam}}$]     & [$\mathrm{\frac{mJy}{beam}}$]   \\
    (1)   & (2) & (3)  & (4) & (5) & (6) &(7)      &   (8)    & (9)  & (10)  & (11)      &   (12)    & (13)  & (14)  \\
    \hline
    12\,GHz & $12.168$ & $11.912 - 12.424$  & $0.69$ & $0.45$ & $-38$ & $1.03$ & $0.031$ & $-28.64$ & $3.38$   & $5458$ & $2$ & $57$ & $4$  \\
    15\,GHz & $15.368$ & $15.112 - 15.624$  & $0.74$ & $0.51$ & $-23$ & $1.09$ & $0.021$ & $-36.07$ & $1.15$   & $7056$ & $5$ & $106$&  $8$ \\
    24\,GHz & $23.568$ & $23.312 - 23.824$  & $0.53$ & $0.35$ & $-22$ & $1.25$ & $0.031$ & $-40.63$ & $1.19$   & $9641$ & $13$  &$225$  &$15$   \\
    43\,GHz & $43.169$ & $42.913 - 43.425$  & $0.21$ & $0.11$ & $-40$ & $1.60$ & $0.016$ & $-42.02$ & $1.73$   & $9550$ & $8$ & $181$&  $7$\\
    \hline
    \end{tabular}
    \\
 \end{center}
    Notes: Columns are;
    (column 1) the band name used in the text, (column 2) the observed central frequency, (column 3) range of observed frequencies, (column 4) the major axis of the nominal beam at each frequency, (column 5) the minor axis of the nominal beam, (column 6) the position angle of the nominal beam, (column 7) the gain corrections discussed in \protect\cref{sec:apriori}, (column 8) scatter of D-terms, (column 9) absolute EVPA correction from VLA observations at $\nu_c$, (column 10) scatter of the model of EVPA, (column 11) the peak Stokes I, (column 12) the noise in Stokes I based on the RMS of the map,
    (column 13) the peak linear polarization, (column 14) the noise in linear polarization calculated using the method outlined in \protect\cref{sec:pol}.
\end{table*}

\subsection{Faraday rotation}
\label{sec:fara}
Following \citet{Sarala2001} we use a circular statistics approach to fitting $\mathrm{RM_{obs}}$ with a fit minimization parameter, $\chi^2_{\mathrm{circ}}$ defined as,
\begin{equation}
    \chi^2_{\mathrm{circ}} = \sum^{n_{\nu}}_{i} \frac{1-\cos \left(2 \left[\mathrm{EVPA} - \mathrm{RM_{obs}} \lambda^2  - \mathrm{EVPA_0}\right]\right)}{1-\cos \left(2 \sigma_{\mathrm{EVPA}}\right)}.
    \label{eqn:circstats}
\end{equation}
Here, $n_{\nu}$ is the number of frequencies we have polarization information for. By minimizing \cref{eqn:circstats} we are able to solve for $\mathrm{EVPA_0}$ \citep{Sarala2001}, 
\begin{equation}
    \mathrm{EVPA_0} = \frac{1}{2} \arctan \left(\frac{\sum_i^n (\sin(E)/[1-\cos (2\sigma_{\mathrm{EVPA}})])}{\sum_i^n (\cos(E)/[1-\cos (2\sigma_{\mathrm{EVPA}})])}\right),
\end{equation}
where $E = 2 (\mathrm{EVPA} - \mathrm{RM_{obs}} \lambda^2 )$. In this formulation we only require $\mathrm{RM_{obs}}$ to model $\mathrm{EVPA}$, without needing to fit the $\mathrm{EVPA_0}$ directly, reducing the number of fitted parameters. 

We model $\mathrm{RM_{obs}}$ for all pixels with a polarized signal greater than $3\sigma_P$ and $5\sigma_{I}$ for at least two frequencies. For pixels with polarization information for more than two frequencies we reject any fits for which the $\chi^2_{\mathrm{circ}}$ of the fit exceeds the 95\% confidence limit. For these pixels, they may have non-$\lambda^2$ dependent Faraday rotation, which occurs when there are complex Faraday rotation effects present \citep{Hovatta2012}.

When we measure $\mathrm{RM_{obs}}$ in the direction of an extra-galactic source we are measuring the combination of different contributions of Faraday rotation such that, 
\begin{equation}
    {\mathrm{RM_{obs}} = \mathrm{RM_{int}}\,(1+z)^{-2} + \mathrm{RM_{IGM}} + \mathrm{RM_{MW}}}.
\end{equation}
Here, $\mathrm{RM_{int}}$ is the intrinsic Faraday rotation of the source which needs to be corrected for the redshift, $z$; $\mathrm{RM_{IGM}}$ and $\mathrm{RM_{MW}}$ are the Intergalactic Medium (IGM) and the Milky Way (MW) contributions to $\mathrm{RM_{obs}}$. The magnitude of $\mathrm{RM_{IGM}}$ is typically between 1 and 10\radm \citep{OSullivan2017} and as such we ignore the contribution of $\mathrm{RM_{IGM}}$. We estimate $\mathrm{RM_{MW}}$ using the $\mathrm{RM}$ catalog of \citet{VanEck2023} taking the median $\mathrm{RM}$ of a $2^{\circ}$ area around the phase center of NRAO\,150 which is $\mathrm{RM_{MW}} = 52$\radm with a standard deviation of $65$\radm.

\subsection{Depolarization}
\label{sec:dep}
Measuring depolarization as a function of frequency can help to determine where the $\mathrm{RM_{int}}$ originates within the source. We model depolarization following the method outlined by \citet{Hovatta2012}, 
\begin{equation}
    \log{\mathrm{m}} = \log{\mathrm{m_0}} + \mathrm{b} \lambda^4.
\end{equation}
Here $\mathrm{m}$ is the measured polarization degree (as a percentage), $\mathrm{m_0}$ is the intrinsic degree of polarization at $\lambda = 0$, and $\mathrm{b}$ is the depolarization in units of $\mathrm{\,m^{-4}}$. 

There are two main causes of depolarization we consider: external and internal Faraday dispersion. If there is magneto-ionic turbulence within an external medium, this causes an EVPA de-coherence across the size of the beam, which results in a reduction in the overall linear polarization fraction. This depolarization increases with the observed wavelength as the size of the resolving beam increases and the variation in $\mathrm{RM_{int}}$ is more prominent for greater $\lambda^2$.

For internal Faraday dispersion, we assume that the polarized emission is co-spatial with the Faraday rotating medium. Emission from the source undergoes differential Faraday rotation depending on where in the medium the emission is generated, which can cause a de-coherence of EVPA and subsequent depolarization. 

Depending on the morphology of the emission region it is also possible to re-align EVPA values, causing "inverse" depolarization \citep{Homan2012}. If we see positive values of $\mathrm{b}$, this indicates the presence of inverse depolarization, which is suggestive of internal Faraday dispersion. The internal Faraday dispersion effect is dependent on the magnitude of $\mathrm{RM}$ such that $\mathrm{b} \sim 2 \mathrm{RM}^2$. If $|\mathrm{b}|/\left(2\,\mathrm{RM}^2\right) \leq 1$, especially when the magnitude of $\mathrm{RM}$ is large, we expect the depolarization observed to be attributed to co-spatial polarized emission and Faraday rotation \citep{Hovatta2012}. If we see large negative values of $\mathrm{b}$ that are larger in magnitude than $2\times\mathrm{RM}^{2}$, this suggests that the depolarization comes from external Faraday dispersion, resulting from random fluctuations in $\mathrm{RM}$ causing EVPA de-coherence.
\section{Results and discussion}
In this section we will discuss the linear polarization, Faraday rotation, $\mathrm{EVPA_0}$, and depolarization properties of NRAO\,150, in order to understand the geometry of its magnetic field.

\subsection{Linear polarization}
In panels {(1)} to {(4)} of \cref{fig:stki} we show the linear polarization intensity maps for the four frequency bands in conjunction with Stokes I contours. From these maps we can see that the linear polarization is present near the VLBI `core' position of the source, with a some log magnitude polarized emission extending outwards towards the jet for the 12 and 15\,GHz maps. The peak polarization is greatest at 24\,GHz, decreasing from $225$\,mJy\,beam$^{-1}$ to $181$\,mJy\,beam$^{-1}$ at 43\,GHz, shown in \cref{tab:calib}. \citet{Molina2014} found for a number of epochs of 22 and 43\,GHz similar linear polarization intensity as shown in panels {(3)} and {(4)}. \citet{Paraschos2024}, shows considerably lower peak polarization intensity at $\sim22$\,GHz than our findings ($\sim 23$\,mJy\,beam$^{-1}$ compared to $225$\,mJy\,beam$^{-1}$). This discrepancy may come from the fact that the source is intrinsically variable (also supported by the finding that the peak Stokes I reported in \citet{Paraschos2024} is 2.5\,Jy\,beam$^{-1}$).

In panels {(1)} to {(4)} of \cref{fig:frac} we show the fractional linear polarization for each frequency. We can see that the fractional linear polarization is greater throughout the 12\,GHz map, with sections of the jet exceeding 20\%, while the fraction near the core is around $\sim$ 5\% . For panels {(2)} to {(4)} we can see that the fractional polarization does not exceed $\sim 10\%$ in the core. We expect polarization of optically thick regions to not exceed 10 -- 15\% and within these regions EVPA runs parallel to the projection of the magnetic field on the plane-of-sky, $\mathbf{B_{\perp}}$ \citep{Pacholczyk1970}. \citet{Molina2014} found fractional polarization of $\approx 5\%, 10\%, 15\%$ for 15, 22, and 43\,GHz observations respectively, and \citet{Molina2016} found a fractional polarization of $\approx 14\%$ around the northern edge of the central region of the source for both frequencies.   

\subsection{Rotation measure}
\begin{figure}
    \centering
    \includegraphics[width=1\columnwidth]{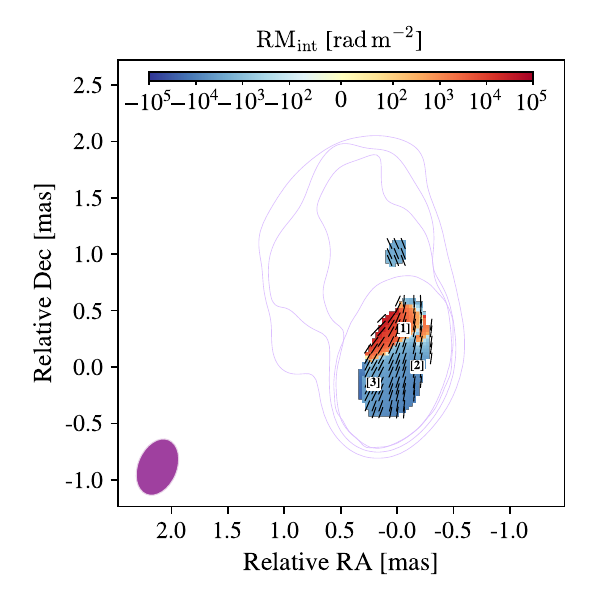}
    \caption{Map of $\mathrm{RM_{int}}$ and $\mathrm{EVPA_0}$. The tick markers show the average $\mathrm{EVPA_0}$ across a $3 \times 3$ pixel region. The contours are the lowest Stokes I contour levels for all frequencies from \protect\cref{fig:stki}. The observed EVPA against $\lambda^2$ for the three pixels marked by numbers 1,2 and 3 are plotted in \cref{fig:rm_pixel}. The purple ellipse in panel shows the restoring beam used for all four frequencies.}
    \label{fig:rm}
\end{figure}

\begin{figure}
    \centering
    \includegraphics[width=1\linewidth]{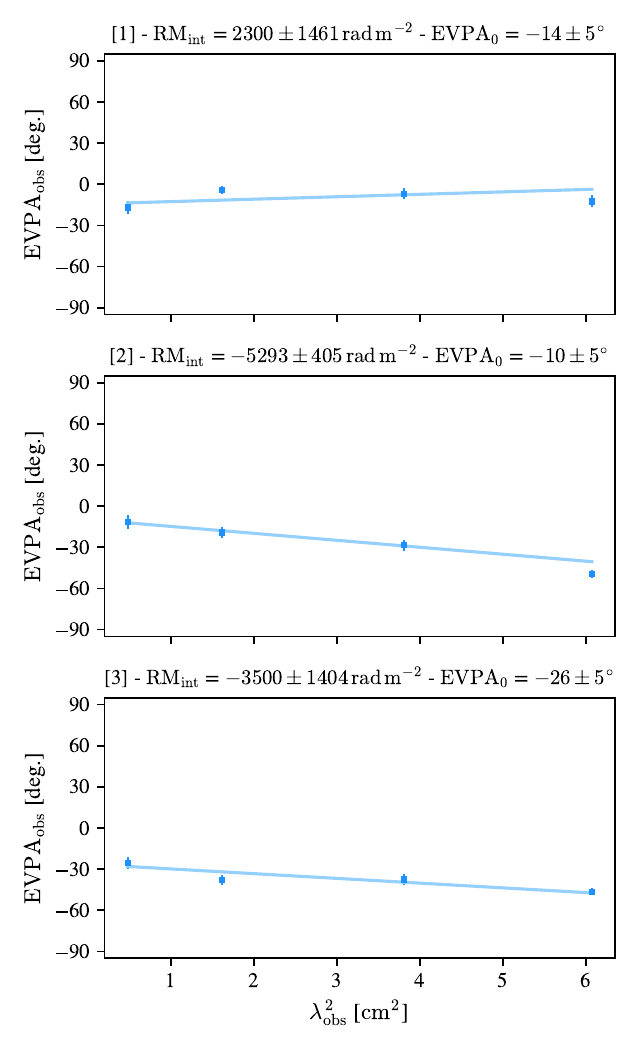}
    \caption{Observed EVPA against $\lambda^2$ for the three pixels marked by numbers 1, 2, and 3 in \cref{fig:rm}. Above each plot is the $\mathrm{RM_{int}}$ (corrected for redshift and foreground contribution as discussed in \cref{sec:fara}) and $\mathrm{EVPA_0}$ for that pixel.}
    \label{fig:rm_pixel}
\end{figure}

In \cref{fig:rm} we show the $\mathrm{RM_{int}}$ map (corrected for the $z$ of NRAO\,150) with ticks showing $\mathrm{EVPA_0}$. Most of the $\mathrm{RM_{int}}$ of NRAO\,150 is negative with the positive peak of ${\mathrm{RM_{int}} = (3.0\pm0.7)\times10^4}$\,\radm and a negative peak of ${\mathrm{RM_{int}} = (-2.0\pm0.7)\times10^4}$\,\radm. Towards the jet there is a gradient in $\mathrm{RM_{int}}$. In \cref{fig:rm_pixel} we show the observed EVPA against $\lambda^2$ for the three pixels marked by numbers 1, 2, and 3 in \cref{fig:rm}. For the pixels the observed EVPA values fit well to the model for $\mathrm{RM_{obs}}$. 


The $\mathrm{EVPA_0}$ shown in \cref{fig:rm} run towards the direction of the jet, gradually bending around the core. We also note that the $\mathrm{EVPA_0}$ values run perpendicular to the gradient in $\mathrm{RM_{int}}$ in the north of the core. In the optically thick regions of a source, we expect $\mathrm{EVPA_0}$ to be aligned with $\mathbf{B_{\perp}}$; the alignment of $\mathrm{EVPA_0}$ is suggestive of a $\mathbf{B_{\perp}}$ pointing in the direction of the jet curving around the core region. Both \citet{Molina2014} and \citet{Molina2016} found a similar circular geometry of EVPA for NRAO\,150, however this was not Faraday rotation corrected.

\subsection{Depolarization and maximum polarization}
\begin{figure}
    \centering
    \includegraphics[width=1\columnwidth]{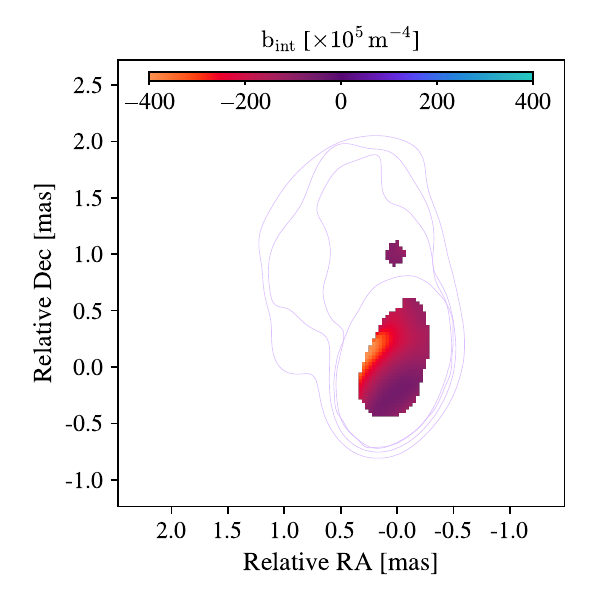}
    \caption{Map of intrinsic depolarization ($\mathrm{b_{int}}$). The contours are the lowest Stokes I contour levels for all frequencies from \protect\cref{fig:stki}.}
    \label{fig:dep}
\end{figure}

In \cref{fig:dep} we show the map of redshift-corrected $\mathrm{b_{int}}$. Here $\mathrm{b_{int}}$ is the depolarization constant $\mathrm{b}$ (discussed in \cref{sec:dep}) scaled by $(1+z)^2$ to account for the redshift of NRAO\,150. Most of the core region has large negative values of $\mathrm{b_{int}}$, which reflects depolarization that increases as a function of increasing observed~$\lambda$.

\begin{figure}
    \centering
    \includegraphics[width=1\columnwidth]{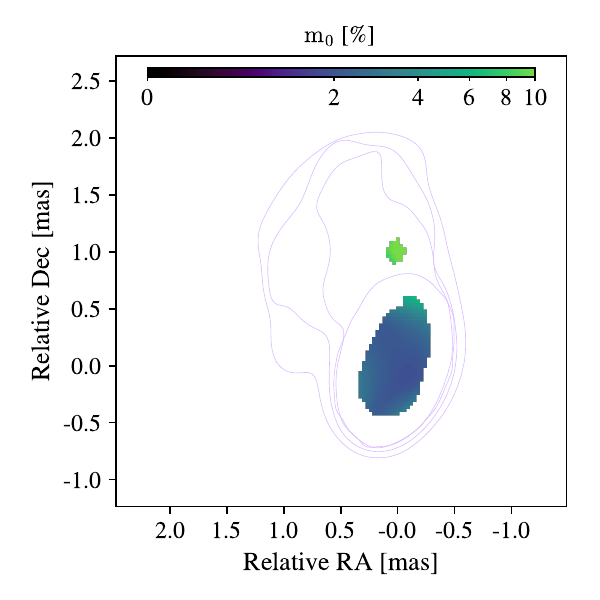}
    \caption{Map of intrinsic polarization degree ($\mathrm{m_{0}}$). The contours are the lowest Stokes I contour levels for all frequencies from \protect\cref{fig:stki}.}
    \label{fig:m0}
\end{figure}

In \cref{fig:m0} we show the intrinsic polarization degree, $\mathrm{m_0}$. Across the entire region, $\mathrm{m_0}$ is around 3\%, with an increase to 6\% on the northern side of the core region. In the small emission we see further down the jet the $\mathrm{m_0}$ increases to $\geq10$\%.  This preferential increase in polarization degree is similar to that seen in the fractional linear polarization maps (\cref{fig:frac}), and the findings of \citet{Molina2014} and \citet{Molina2016}. Values of $\mathrm{m_0}$ around a few percent hint that the emission we are probing is likely optically thick for which we expect lower fractional polarization \citep{Pacholczyk1970}.

\begin{figure}
    \centering
    \includegraphics[width=1\columnwidth]{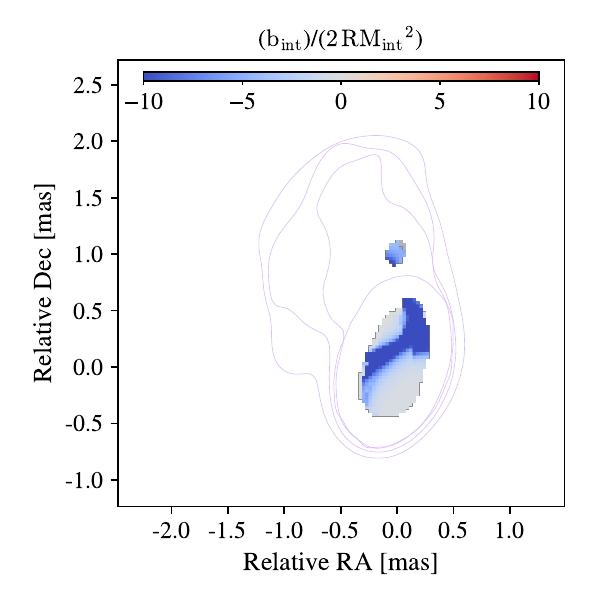}
    \caption{Map of the ratio of $\mathrm{b_{int}}/\left(2\,\mathrm{RM_{int}}^2\right)$. Grey pixels show rejected fits of $\mathrm{RM}$. The contours are the lowest Stokes I contour levels for all frequencies from \protect\cref{fig:stki}.}
    \label{fig:ratio}
\end{figure}

In the map of ratios of $\mathrm{b_{int}}/\left(2\,\mathrm{RM_{int}}^2\right)$ shown in \cref{fig:ratio}, 31\% of pixels have ratios greater than --1, indicating that the majority of pixels show $\mathrm{b_{int}}/\left(2\,\mathrm{RM_{int}}^2\right) < -1$. The pixels with negative ratio values are spatially close to the point at which the gradient seen in \cref{fig:rm} passes through zero. We can see that the ratio of $\mathrm{b_{int}}/\left(2\,\mathrm{RM_{int}}^2\right)$ decreases in magnitude away from this transition going towards $\mathrm{b_{int}}/\left(2\,\mathrm{RM_{int}}^2\right) > -1$. Ratios $\geq -1$ indicate that the depolarization we see is explainable by the magnitude of $\mathrm{RM}$ causing internal Faraday rotation, without needing to invoke magneto-ionic turbulence as is required in the external Faraday dispersion case. 

\subsection{Faraday rotation gradients}
\begin{figure}
    \centering
    \includegraphics[width=1\columnwidth]{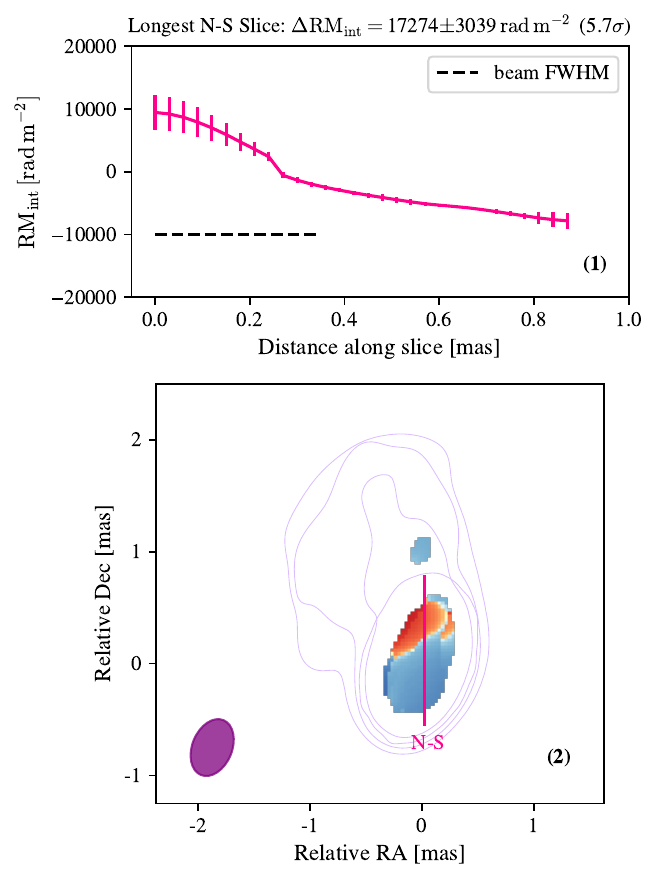}
    \caption{Slice of $\mathrm{RM_{int}}$ of NRAO\,150 of the longest North-South (N-S) slice {(1)}. Panel {(2)} is an illustration of slice locations on RM map from \protect\cref{fig:rm}. The black dashed line shows the beam full-width half maximum (FWHM) in the direction of the slice.}
    \label{fig:rmgrad}
\end{figure}

In \cref{fig:rmgrad} we show the longest North-South (N-S), panel {(1)} shown in panel {(2)}. \citet{Taylor2010} first proposed a list of criteria for Faraday rotation gradients to be considered significant, which was followed up by \citet{Hovatta2012} who extensively tested the validity of $\mathrm{RM_{int}}$ gradients, finding that if a gradient has a $\delta \mathrm{RM}$ with $\geq 3 \sigma$ confidence and is $\geq 1.4\times$ the beam full-width half-maximum (FWHM) in the direction of the gradient it is significant. The gradient shown in \cref{fig:rmgrad} has a confidence in $\delta \mathrm{RM_{int}}$ of $5.7 \sigma$ and and is $2\times$ the beam FWHM in the direction of the slice at 0.9\,mas in length corresponding to a projected size of $\sim 7.7$ pc.

\citet{Broderick2010} conducted parsec-scale general relativistic magnetohydrodynamic simulations of AGN jets testing the expected morphology of Faraday rotation. One of their primary findings was that the presence of a significant Faraday rotation gradient indicates the presence of a helically/toroidal magnetic field located within the jet or core of the AGN. This is because foreground clouds, while contributing the the overall Faraday rotation of some AGN, should not show Faraday rotation gradients that are correlated with jet structures \citep{Broderick2010}.

In their study of Faraday rotation gradients, \citet{Gabuzda2015} found a gradient across the core region of NRAO\,150 (observed in 2001 between $8.1 - 15.2$\,GHz) with a significance of $4.2\sigma$ ($\Delta \mathrm{RM_{int}} = 3944\pm934$ \radm) with a length of 1.5\,mas. Following the findings of \citet{Broderick2010}, \citet{Gabuzda2015} concluded that the presence of a significant gradient in the core of NRAO\,150 is indicative of the presence of a helical/toroidal magnetic field within the innermost jet of the AGN, even if the region is not optically thin, given the significance and continuity of the RM gradients seen for NRAO\,150. We see similar significant and continuous gradients across the core region of NRAO\,150. We conclude that the Faraday rotation gradients we see, especially our longest Faraday rotation gradient with $\Delta \mathrm{RM_{int}} = (15954\pm2661)$ \radm across a angular distance of 0.9\,mas ($\sim 7.7$ pc), indicate a helical/toroidal magnetic field is present within the innermost jet of NRAO\,150.


\subsection{The magnetic field strength in NRAO\,150}
In \cref{fig:rmgrad} we see that $\mathrm{RM_{int}}$ varies across the source, with a gradient in $\mathrm{RM_{int}}$ of $\Delta \mathrm{RM_{int}} = (17274\pm3039)$ \radm with a maximum magnitude value of $|\mathrm{RM_{int}}| = (9453\pm2737)$ \radm. This occurs over an apparent size scale of $\sim7.7$ pc. Assuming an apparent opening angle of $19.4^{\circ}$ \citep{Pushkarev2017} this results in an apparent distance from the jet base of $\sim 23$ pc across the gradient. If we further assume that at 1 pc from the core the free electron number density is between $10^2$ and $10^4$ $\mathrm{cm^{-3}}$ \citep{Lobanov1998}, the magnetic field strength and electron number density both go as $\sim r^{-1}$ \citep{Lobanov1998,Pushkarev2012,Nokhrina2024}, and we are integrating along a path-length of 22 pc, the relationship between $\mathrm{RM_{int}}$ and $B_{||}$ from \cref{eqn:RM} becomes,
\begin{equation}
    B_{||,r=1} \sim 13 \left(\frac{\mathrm{RM_{int}}}{1000\mathrm{\,rad\,m^{-2}}}\right) \left(\frac{n_{r=1}}{100\mathrm{\,cm^{-3}}}\right)^{-1} \mu \mathrm{G},
    \label{eqn:B}
\end{equation}
where $n_{r=1}$ is the electron density at 1 pc from the core. Setting $\mathrm{RM_{int}} = (9453\pm2737)$ \radm we get a lower limit of $B_{||,0} \sim 1.2\,\mu$G (assuming $n_{0} = 10^4$ $\mathrm{cm^{-3}}$) and an upper limit of $B_{||,0} \sim 122\mu$G ($n_{0} = 10^2$ $\mathrm{cm^{-3}}$). This is similar to the LOS magnetic field strength estimated by \citet{Hovatta2019} for 3C\,273 of $62 \,\mu$G assuming an electron number density of $1000\,\mathrm{cm^{-3}}$ coming from the narrow line region \citep{Zavala2004}. 

\citet{Nokhrina2024} using a relationship between core-shift \citep{Lobanov1998} and the energy equipartition assumption found that the typical total magnetic field strength at 1 pc from the core of AGN is $\sim 84$ mG. If we instead assume that the LOS magnetic field strength and the plane-of-sky magnetic field strength are the same, i.e. $B_{||,r=1} \sim 59$ mG, this results in an electron number density of $n_{r=1} \sim 2\times10^{-1}\,\mathrm{cm^{-3}}$. To differentiate between the scenarios of a weak magnetic field ($1.2 - 122\mu$G) and a small electron number density we plan to conduct simulations in a future publication similar to those done by \cite{MacDonald2021}.
\section{Summary and conclusions}

In this work we presented multi-frequency polarimetric observations of quasar NRAO\,150 taken using the VLBA+EF telescopes at 12, 15, 24, and 43\,GHz. This work builds on previous studies of the source, with the unique quality of having high-frequency (between 12 -- 43\,GHz) polarimetric observations that are quasi-simultaneous. From the linearly polarized emission, we calculated the Faraday rotation, intrinsic electric vector position angle ($\mathrm{EVPA_0}$), and depolarization properties of the source.

We find a circular pattern of $\mathrm{EVPA_0}$ and the presence of Faraday rotation gradients across the core region of NRAO\,150, both of which are consistent with a helical/toroidal magnetic field. The predicted maximum polarization fraction, $\mathrm{m_0}$, for this region is $\approx 3\%$, consistent with emission from optically thick regions in AGN. Together with the magnitude of Faraday rotation in this region, we conclude that the observed polarized emission is produced within the jet, opposed to an external medium. 

The results of this study, in conjunction with previous findings from \citet{Molina2014}, \citet{Gabuzda2015}, and \citet{Molina2016}, show there is a helical/toroidal magnetic field within the innermost jet region of NRAO\,150.

\begin{acknowledgements}
      We thank Sebastiano D. von Fellenberg for his helpful comments on the manuscript draft and we thank the anonymized reviewer for their constructive feedback during the review process. M2FINDERS project has received funding from the European Research Council (ERC) under the European Union’s Horizon 2020 research and innovation programme (grant agreement No 101018682). MW is supported by a~Ramón y Cajal grant RYC2023-042988-I from the Spanish Ministry of Science and Innovation. The VLBA is an instrument of the National Radio Astronomy Observatory. The National Radio Astronomy Observatory is a facility of the National Science Foundation operated by Associated Universities, Inc. This work made use of the Swinburne University of Technology software correlator \citep{Deller2011}, developed as part of the Australian Major National Research Facilities Programme and operated under licence. This research has made use of data from the MOJAVE database that is maintained by the MOJAVE team \citep{Lister2018}. This study makes use of VLBA data from the VLBA-BU Blazar Monitoring Program (BEAM-ME and VLBA-BU-BLAZAR; \href{http://www.bu.edu/blazars/BEAM-ME.html}{http://www.bu.edu/blazars/BEAM-ME.html}), funded by NASA through the Fermi Guest Investigator Program. YYK was supported by the MuSES project which has received funding from the European Research Council (ERC) under the European Union's Horizon 2020 Research and Innovation Programme (grant agreement No 101142396).

\end{acknowledgements}

\section*{Data Availability}
The maps of $\mathrm{RM_{obs}}$ and $\mathrm{EVPA_0}$ shown in \cref{fig:rm} and associated uncertainties are only available in electronic form at the CDS via anonymous ftp to \href{http://cdsarc.u-strasbg.fr/}{cdsarc.u-strasbg.fr} (130.79.128.5) or via \href{http://cdsweb.u-strasbg.fr/cgi-bin/qcat?J/A+A/}{http://cdsweb.u-strasbg.fr/cgi-bin/qcat?J/A+A/}.

\section*{Appendix: Alternative polarization calibration using \texttt{CASA/PolSolve}}
\begin{figure}
    \centering
    \includegraphics[width=1\columnwidth]{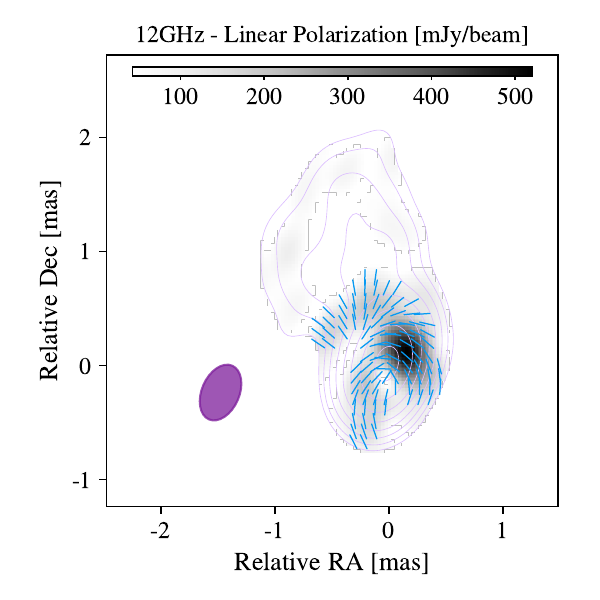}
    \caption{Stokes I contours and linear polarization intensity (shown in gray-scale) of NRAO\,150 for the 12\,GHz band derived using a-priori calibration from rPicard/CASA, polarization leakage calibration from \texttt{PolSolve}, and absolute EVPA correction using CASA.}
    \label{app:12GHz}
\end{figure}

As part of this work, the rPicard/CASA calibrated data (discussed in \cref{sec:apriori}) was also polarization leakage calibrated using \texttt{PolSolve v2}\footnote{\href{https://github.com/marti-vidal-i/casa-poltools}{https://github.com/marti-vidal-i/casa-poltools}} \citep{MartiVidal2021} and absolute EVPA corrected using CASA function \texttt{polcal}. After D-term and absolute EVPA correction the 15\,GHz, 24\,GHz, and 43\,GHz linear polarization and EVPA maps were similar to those seen in \cref{fig:stki}, however we saw significant differences for the 12\,GHz band. In \cref{app:12GHz} we show the 12\,GHz linear polarization and EVPA distribution using these alternative methods. Compared to what we see for the 12\,GHz band in \cref{fig:stki}, we can see a flip in EVPA values of $\sim 90^\circ$ in the core region and a significant difference in overall linear polarization morphology. We also see a much larger polarized intensity with a peak of $520$\,mJy\,beam$^{-1}$. 

The D-terms derived by \texttt{PolSolve} took 16 iterations to converge for the 12\,GHz band, whereas it took 2, 5, and 4 iterations for the 15\,GHz, 24\,GHz, and 43\,GHz bands, indicating that the polarization leakage was not well characterized for the 12\,GHz band. The significant differences in linear polarization of the 12\,GHz band using these CASA based methods may be due to a bug we found in the rPicard/CASA pipeline, which flips Stokes Q and U, resulting in a rotation of the EVPA distribution by $45^{\circ}$, an issue with D-term characterization using \texttt{PolSolve}, or the application of CASA function \texttt{polcal} in absolute EVPA calibration for VLBI data. Due to this reason we elected to use GPCAL and AIPS function \text{CLCOR} to correct for polarization leakage and perform absolute EVPA correction.

%
%

\bibliographystyle{aa} 
\bibliography{aanda} 

\begin{thebibliography}{34}
\expandafter\ifx\csname natexlab\endcsname\relax\def\natexlab#1{#1}\fi

\bibitem[{{Acosta-Pulido} {et~al.}(2010){Acosta-Pulido}, {Agudo}, {Barrena}, {Ramos Almeida}, {Manchado}, \& {Rodr{\'\i}guez-Gil}}]{AcostaPulido2010}
{Acosta-Pulido}, J.~A., {Agudo}, I., {Barrena}, R., {et~al.} 2010, \aap, 519, A5

\bibitem[{{Agudo} {et~al.}(2007){Agudo}, {Bach}, {Krichbaum}, {Marscher}, {Gonidakis}, {Diamond}, {Perucho}, {Alef}, {Graham}, {Witzel}, {Zensus}, {Bremer}, {Acosta-Pulido}, \& {Barrena}}]{Agudo2007}
{Agudo}, I., {Bach}, U., {Krichbaum}, T.~P., {et~al.} 2007, \aap, 476, L17

\bibitem[{Broderick \& McKinney(2010)}]{Broderick2010}
Broderick, A.~E. \& McKinney, J.~C. 2010, The Astrophysical Journal, 725, 750

\bibitem[{{Deller} {et~al.}(2011){Deller}, {Brisken}, {Phillips}, {Morgan}, {Alef}, {Cappallo}, {Middelberg}, {Romney}, {Rottmann}, {Tingay}, \& {Wayth}}]{Deller2011}
{Deller}, A.~T., {Brisken}, W.~F., {Phillips}, C.~J., {et~al.} 2011, \pasp, 123, 275

\bibitem[{{Ferri{\`e}re} {et~al.}(2021){Ferri{\`e}re}, {West}, \& {Jaffe}}]{Ferriere2021}
{Ferri{\`e}re}, K., {West}, J.~L., \& {Jaffe}, T.~R. 2021, \mnras, 507, 4968

\bibitem[{Gabuzda {et~al.}(2015)Gabuzda, Knuettel, \& Reardon}]{Gabuzda2015}
Gabuzda, D.~C., Knuettel, S., \& Reardon, B. 2015, Monthly Notices of the Royal Astronomical Society, 450, 2441–2450

\bibitem[{{Homan}(2012)}]{Homan2012}
{Homan}, D.~C. 2012, \apjl, 747, L24

\bibitem[{{Homan} {et~al.}(2021){Homan}, {Cohen}, {Hovatta}, {Kellermann}, {Kovalev}, {Lister}, {Popkov}, {Pushkarev}, {Ros}, \& {Savolainen}}]{Homan2021}
{Homan}, D.~C., {Cohen}, M.~H., {Hovatta}, T., {et~al.} 2021, \apj, 923, 67

\bibitem[{{Homan} {et~al.}(2024){Homan}, {Roth}, \& {Pushkarev}}]{Homan2024}
{Homan}, D.~C., {Roth}, J.~S., \& {Pushkarev}, A.~B. 2024, \aj, 167, 11

\bibitem[{{Hovatta} {et~al.}(2012){Hovatta}, {Lister}, {Aller}, {Aller}, {Homan}, {Kovalev}, {Pushkarev}, \& {Savolainen}}]{Hovatta2012}
{Hovatta}, T., {Lister}, M.~L., {Aller}, M.~F., {et~al.} 2012, \aj, 144, 105

\bibitem[{{Hovatta} {et~al.}(2019){Hovatta}, {O'Sullivan}, {Mart{\'\i}-Vidal}, {Savolainen}, \& {Tchekhovskoy}}]{Hovatta2019}
{Hovatta}, T., {O'Sullivan}, S., {Mart{\'\i}-Vidal}, I., {Savolainen}, T., \& {Tchekhovskoy}, A. 2019, \aap, 623, A111

\bibitem[{{Hovatta} {et~al.}(2009){Hovatta}, {Valtaoja}, {Tornikoski}, \& {L{\"a}hteenm{\"a}ki}}]{Hovatta2009}
{Hovatta}, T., {Valtaoja}, E., {Tornikoski}, M., \& {L{\"a}hteenm{\"a}ki}, A. 2009, \aap, 494, 527

\bibitem[{{Janssen} {et~al.}(2019){Janssen}, {Goddi}, {van Bemmel}, {Kettenis}, {Small}, {Liuzzo}, {Rygl}, {Mart{\'\i}-Vidal}, {Blackburn}, {Wielgus}, \& {Falcke}}]{Janssen2019}
{Janssen}, M., {Goddi}, C., {van Bemmel}, I.~M., {et~al.} 2019, \aap, 626, A75

\bibitem[{{Jorstad} \& {Marscher}(2016)}]{Jorstad2016}
{Jorstad}, S. \& {Marscher}, A. 2016, Galaxies, 4, 47

\bibitem[{{Komatsu} {et~al.}(2009){Komatsu}, {Dunkley}, {Nolta}, {Bennett}, {Gold}, {Hinshaw}, {Jarosik}, {Larson}, {Limon}, {Page}, {Spergel}, {Halpern}, {Hill}, {Kogut}, {Meyer}, {Tucker}, {Weiland}, {Wollack}, \& {Wright}}]{Komatsu2009}
{Komatsu}, E., {Dunkley}, J., {Nolta}, M.~R., {et~al.} 2009, \apjs, 180, 330

\bibitem[{{Lister} {et~al.}(2018){Lister}, {Aller}, {Aller}, {Hodge}, {Homan}, {Kovalev}, {Pushkarev}, \& {Savolainen}}]{Lister2018}
{Lister}, M.~L., {Aller}, M.~F., {Aller}, H.~D., {et~al.} 2018, \apjs, 234, 12

\bibitem[{{Lobanov}(1998)}]{Lobanov1998}
{Lobanov}, A.~P. 1998, \aap, 330, 79

\bibitem[{{MacDonald} \& {Nishikawa}(2021)}]{MacDonald2021}
{MacDonald}, N.~R. \& {Nishikawa}, K.~I. 2021, \aap, 653, A10

\bibitem[{{Mart{\'\i}-Vidal} {et~al.}(2021){Mart{\'\i}-Vidal}, {Mus}, {Janssen}, {de Vicente}, \& {Gonz{\'a}lez}}]{MartiVidal2021}
{Mart{\'\i}-Vidal}, I., {Mus}, A., {Janssen}, M., {de Vicente}, P., \& {Gonz{\'a}lez}, J. 2021, \aap, 646, A52

\bibitem[{{Molina} {et~al.}(2014){Molina}, {Agudo}, {G{\'o}mez}, {Krichbaum}, {Mart{\'\i}-Vidal}, \& {Roy}}]{Molina2014}
{Molina}, S.~N., {Agudo}, I., {G{\'o}mez}, J.~L., {et~al.} 2014, \aap, 566, A26

\bibitem[{Molina {et~al.}(2016)Molina, Agudo, Gómez, Krichbaum, Martí-Vidal, \& Roy}]{Molina2016}
Molina, S.~N., Agudo, I., Gómez, J.~L., {et~al.} 2016, Galaxies, 4

\bibitem[{{Nokhrina} \& {Pushkarev}(2024)}]{Nokhrina2024}
{Nokhrina}, E.~E. \& {Pushkarev}, A.~B. 2024, \mnras, 528, 2523

\bibitem[{{O'Sullivan} {et~al.}(2017){O'Sullivan}, {Purcell}, {Anderson}, {Farnes}, {Sun}, \& {Gaensler}}]{OSullivan2017}
{O'Sullivan}, S.~P., {Purcell}, C.~R., {Anderson}, C.~S., {et~al.} 2017, \mnras, 469, 4034

\bibitem[{Pacholczyk(1970)}]{Pacholczyk1970}
Pacholczyk, A. 1970, Radio Astrophysics: Nonthermal Processes in Galactic and Extragalactic Sources, Astronomy and Astrophysics Series (W. H. Freeman)

\bibitem[{{Paraschos} {et~al.}(2024){Paraschos}, {Debbrecht}, {Kramer}, {Traianou}, {Liodakis}, {Krichbaum}, {Kim}, {Janssen}, {Nair}, {Savolainen}, {Ros}, {Bach}, {Hodgson}, {Lisakov}, {MacDonald}, \& {Zensus}}]{Paraschos2024}
{Paraschos}, G.~F., {Debbrecht}, L.~C., {Kramer}, J.~A., {et~al.} 2024, \aap, 686, L5

\bibitem[{{Park} {et~al.}(2021){Park}, {Byun}, {Asada}, \& {Yun}}]{Park2021}
{Park}, J., {Byun}, D.-Y., {Asada}, K., \& {Yun}, Y. 2021, \apj, 906, 85

\bibitem[{{Pushkarev} {et~al.}(2012){Pushkarev}, {Hovatta}, {Kovalev}, {Lister}, {Lobanov}, {Savolainen}, \& {Zensus}}]{Pushkarev2012}
{Pushkarev}, A.~B., {Hovatta}, T., {Kovalev}, Y.~Y., {et~al.} 2012, \aap, 545, A113

\bibitem[{{Pushkarev} {et~al.}(2017){Pushkarev}, {Kovalev}, {Lister}, \& {Savolainen}}]{Pushkarev2017}
{Pushkarev}, A.~B., {Kovalev}, Y.~Y., {Lister}, M.~L., \& {Savolainen}, T. 2017, \mnras, 468, 4992

\bibitem[{{Richards} {et~al.}(2011){Richards}, {Max-Moerbeck}, {Pavlidou}, {King}, {Pearson}, {Readhead}, {Reeves}, {Shepherd}, {Stevenson}, {Weintraub}, {Fuhrmann}, {Angelakis}, {Zensus}, {Healey}, {Romani}, {Shaw}, {Grainge}, {Birkinshaw}, {Lancaster}, {Worrall}, {Taylor}, {Cotter}, \& {Bustos}}]{Richards2011}
{Richards}, J.~L., {Max-Moerbeck}, W., {Pavlidou}, V., {et~al.} 2011, \apjs, 194, 29

\bibitem[{{Sarala} \& {Jain}(2001)}]{Sarala2001}
{Sarala}, S. \& {Jain}, P. 2001, \mnras, 328, 623

\bibitem[{{Shepherd}(1997)}]{Shepherd1997}
{Shepherd}, M.~C. 1997, in Astronomical Society of the Pacific Conference Series, Vol. 125, Astronomical Data Analysis Software and Systems VI, ed. G.~{Hunt} \& H.~{Payne}, 77

\bibitem[{{Taylor} \& {Zavala}(2010)}]{Taylor2010}
{Taylor}, G.~B. \& {Zavala}, R. 2010, \apjl, 722, L183

\bibitem[{{Van Eck} {et~al.}(2023){Van Eck}, {Gaensler}, {Hutschenreuter}, {Livingston}, {Ma}, {Riseley}, {Thomson}, {Adebahr}, {Basu}, {Birkinshaw}, {En{\ss}lin}, {Heald}, {Mao}, \& {McClure-Griffiths}}]{VanEck2023}
{Van Eck}, C.~L., {Gaensler}, B.~M., {Hutschenreuter}, S., {et~al.} 2023, \apjs, 267, 28

\bibitem[{{Zavala} \& {Taylor}(2004)}]{Zavala2004}
{Zavala}, R.~T. \& {Taylor}, G.~B. 2004, \apj, 612, 749

\end{thebibliography}

\end{document}